\documentstyle[aps,preprint,psfig,graphicx]{revtex}

\draft

\begin{document}

\title{Properties of $\beta$-stable neutron star
       matter with hyperons}

\author{I.\ Vida\~na, A.\ Polls, and A.\ Ramos}

\address{Departament d'Estructura i Constituents de la Mat\`eria,
         Universitat de Barcelona, E-08028 Barcelona, Spain}

\author{\O.\ Elgar\o y, L.\ Engvik, and M.\  Hjorth-Jensen}

\address{Department of Physics, University of Oslo, N-0316 Oslo, Norway}

\maketitle

\begin{abstract}

We present results from many-body calculations
for $\beta$-stable neutron star
matter with nucleonic and
hyperonic degrees of freedom, employing the most recent parametrizations
of the baryon-baryon interaction of the Nijmegen group.
It is found that the only strange baryons emerging in $\beta$-stable matter
up to total baryonic densities of 1.2 fm$^{-3}$ are $\Sigma^-$ and
$\Lambda$.
The corresponding equations of state are thence used to compute properties 
of neutron stars such as the masses, moments of inertia and radii.
We also study the possibility of forming a hyperon superfluid and discuss
its implications for neutron stars.

\end{abstract}



\section{Introduction}

The physics of compact objects like neutron stars offers
an intriguing interplay between nuclear processes  and
astrophysical observables.
Neutron stars exhibit conditions far from those
encountered on earth; typically, expected densities $\rho$
of a neutron star interior are of the
order of $10^3$ or more times the density 
$\rho_d\approx 4\cdot 10^{11}$ g/cm$^{3}$ at 'neutron drip',
the density at which nuclei begin to
dissolve and merge together.
Thus, the determination of an equation of state (EoS)
for dense matter is essential to calculations of neutron
star properties. The EoS determines properties  such as
the mass range, the mass-radius relationship, the crust
thickness and the cooling rate.
The same EoS is also crucial
in calculating the energy released in a supernova explosion.

At densities near to the saturation density of nuclear
matter, ( with number density $n_0=0.16$ fm$^{-3}$), 
we expect the matter to be composed of mainly neutrons, protons 
and electrons in $\beta$-equilibrium, since neutrinos have on average a
mean free path larger than the radius of the neutron star. The
equilibrium conditions can then be summarized as
\begin{equation}
    \mu_n=\mu_p+\mu_e,  \hspace{1cm} n_p = n_e,
     \label{eq:npebetaequilibrium}
\end{equation}
where $\mu_i$ and $n_i$ refer to the chemical potential and number density
in fm$^{-3}$ of particle species $i$, respectively.
At the saturation density of nuclear matter, $n_0$,
the electron chemical potential is
of the order $\sim 100$ MeV.
Once the rest mass of the muon is exceeded, it becomes
energetically favorable for an electron at the top
of the $e^-$ Fermi surface to decay into a
$\mu^-$. We then develop a Fermi sea of degenerate negative muons,
and we have to modify the charge balance according to $n_p = n_e+n_{\mu}$,
and require that $\mu_e = \mu_{\mu}$.

As the density increases, new hadronic degrees of freedom may 
appear in addition
to neutrons and protons.
One such degree of freedom is hyperons, baryons with a
strangeness content.
Contrary to terrestrial conditions where hyperons are unstable and decay
into nucleons through the weak interaction, the equilibrium conditions
in neutron stars can make the inverse process happen, so that the
formation of hyperons becomes energetically favorable.     
As soon as the chemical potential
of the neutron becomes sufficiently large, energetic neutrons
can decay via weak strangeness non-conserving interactions
into $\Lambda$ hyperons leading to a $\Lambda$ Fermi sea
with $\mu_{\Lambda}=\mu_n$.
However, one expects $\Sigma^-$ to appear via
\begin{equation}
    e^-+n \rightarrow \Sigma^- +\nu_e,
\end{equation}
at lower densities than the $\Lambda$, even though $\Sigma^-$ is more
massive. The negatively charged hyperons
appear in the ground state of matter when their masses
equal $\mu_e+\mu_n$, while the neutral hyperon $\Lambda$
appears when $\mu_n$ equals its mass. Since the
electron chemical potential in matter is larger than
the mass difference $m_{\Sigma^-}-m_{\Lambda}= 81.76$ MeV,
$\Sigma^-$ will appear at lower densities than $\Lambda$.
For matter with hyperons as well
the chemical equilibrium condition becomes,
\begin{eqnarray}
    \mu_{\Xi^-}=\mu_{\Sigma^-} = \mu_n + \mu_e, \nonumber \\
    \mu_{\Lambda} = \mu_{\Xi^0}=\mu_{\Sigma^0} = \mu_n , \nonumber \\
    \mu_{\Sigma^+} = \mu_p = \mu_n - \mu_e .
    \label{eq:beta_baryonicmatter}
\end{eqnarray}
We have omitted isobars $\Delta$, see the discussion below.

Hyperonic degrees of freedom have been considered by several authors,
but mainly within the framework of relativistic
mean field models \cite{prakash97,pke95,ms96} or parametrized
effective interactions \cite{bg97}, 
see also Balberg {\em et al.} \cite{blc99}
for a recent update. Realistic hyperon-nucleon interactions
were employed by Schulze {\em et al.}  recently, see Ref.\ \cite{bbs98},
in a many-body calculation in order to study
where hyperons appear in neutron star matter. 
All these works show that hyperons appear at densities of the order of
$\sim 2n_0$.

In Ref.\ \cite{bbs98} however,
one was only able to fix the density where $\Sigma^-$ appears, since 
only a hyperon-nucleon interaction was employed. As soon as
$\Sigma^-$ appears, one needs a hyperon-hyperon interaction in order
to estimate e.g., the self-energy of $\Lambda$. 
The aim of this work is thus 
to present results from
many-body calculations of hyperonic degrees of freedom
for $\beta$-stable neutron star matter employing interactions
which also account for strangeness $S < -1$.
To achieve this goal,
our many-body scheme starts with the most recent
parametrization
of the free baryon-baryon potentials
for the complete  baryon octet
as defined by Stoks and Rijken in Ref.\
\cite{sr99}. 
This entails a microscopic
description of matter starting from
realistic nucleon-nucleon, hyperon-nucleon
and hyperon-hyperon interactions.
In a recent work \cite{isaac99}
we have developed a formalism for microscopic
Brueckner-type calculations of dense nuclear
matter that includes all types of
baryon-baryon interactions and allows to treat
any asymmetry on the fractions of the
different species ($n, p, \Lambda, \Sigma^-, \Sigma^0, \Sigma^+,
\Xi^-$ and $\Xi^0$).  Results for various fractions of the above
particles were also discussed.

Here we extend the calculations of Ref.\ \cite{isaac99}
to studies of $\beta$-stable neutron star matter.
Our results, together with a brief summary of the formalism
discussed in Ref.\ \cite{isaac99}, are presented  
in section \ref{sec:sec2}. There we discuss the
equation of state (EoS) and the composition of $\beta$-stable matter with
various baryon-baryon potentials. Based on the composition of
matter we present also results for baryon superfluidity and discuss the
possible neutron star structures.

\section{Equation of state and composition of $\beta$-stable matter}
\label{sec:sec2}

Our many-body scheme starts with the most recent
parametrization
of the free baryon-baryon potentials
for the complete  baryon octet
as defined by Stoks and Rijken in Ref.\
\cite{sr99}.
This potential model, which aims at describing all
interaction channels
with strangeness from $S=0$ to $S=-4$,
is based on SU(3) extensions
of the Nijmegen potential models \cite{rsy98}
for the $S=0$ and $S=-1$ channels, which
are fitted to the available body of experimental
data and constrain all free parameters in the model.
In our discussion we employ
the interaction version NSC97e of Ref.\ \cite{sr99}, since this
model, together with the model NSC97f of Ref.\ \cite{sr99}, result in
the best predicitions for hypernuclear observables \cite{rsy98}.
For a discussion of other interaction models, see Refs.\ 
\cite{sr99,sl99}.

\subsection{Formalism}

\begin{figure}[hbtp]
   \includegraphics[totalheight=20cm,angle=0,bb= 0 -10 350 730]{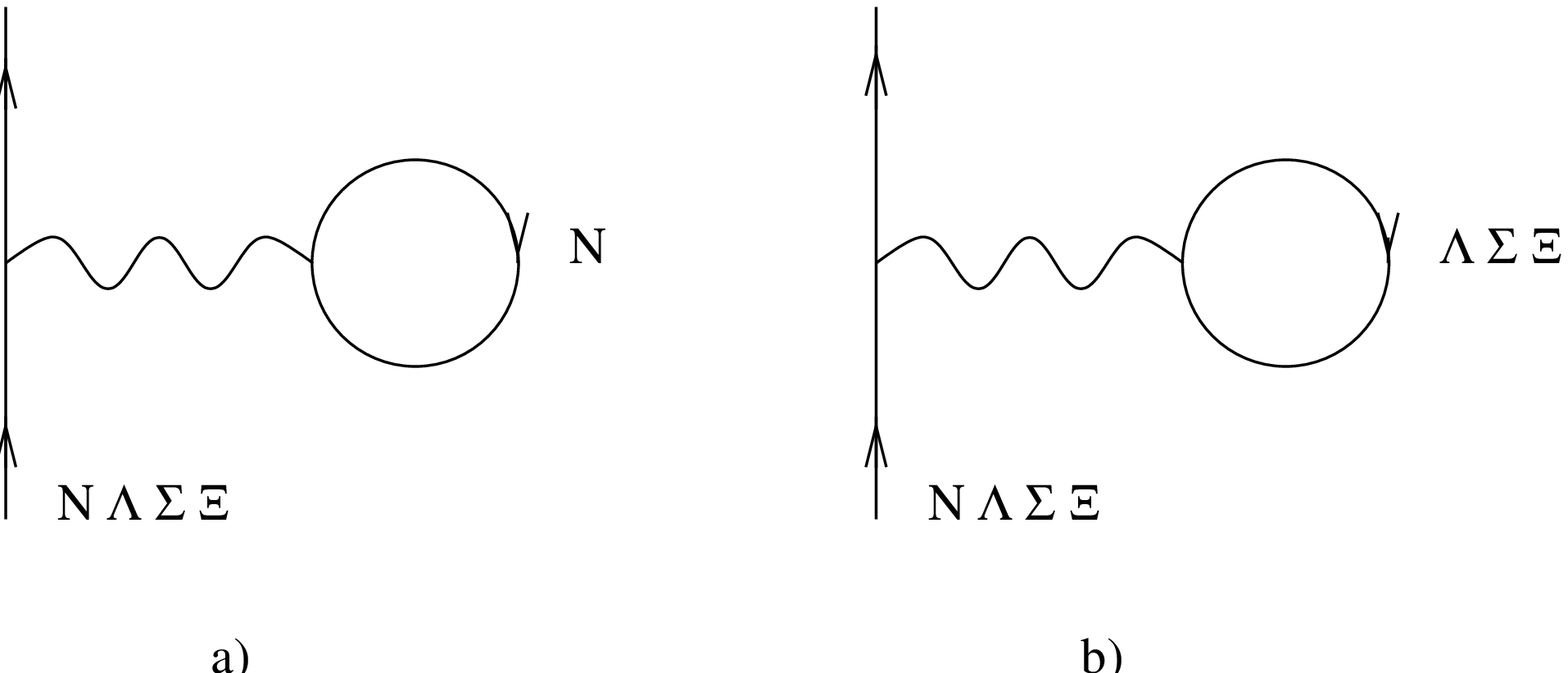}
   \caption{Goldstone diagrams for the single-particle potential $u$.
            a) represents the contribution from nucleons only as hole
            states while b) includes only hyperons as hole states.
            The wavy line represents the $G$-matrix.}
   \label{fig:upot}
\end{figure}

With a given interaction model,
the next step is to introduce effects from the nuclear medium.
Here we will construct the so-called $G$-matrix, which
takes into account short-range correlations for all strangeness
sectors, and solve the equations for the single-particle energies
of the various baryons self-consistently.
The $G$-matrix is formally given by
\begin{eqnarray}
   \left\langle B_1B_2\right |G(\omega)\left | B_3B_4 \right\rangle=
   \left\langle B_1B_2\right |V\left | B_3B_4 \right\rangle+&\nonumber\\
   \sum_{B_5B_6}\left\langle B_1B_2\right |V\left | B_5B_6 \right\rangle
   \frac{1}{\omega-\varepsilon_{B_5}-\varepsilon_{B_6}+ \imath\eta}&\nonumber\\
   \times\left\langle B_5B_6\right |G(\omega)\left | B_3B_4 \right\rangle&.
   \label{eq:gmatrix}
\end{eqnarray}
Here $B_i$ represents all possible baryons $n$, $p$, $\Lambda$, $\Sigma^{-}$,
$\Sigma^0$, $\Sigma^+$, $\Xi^-$ and $\Xi^0$ and their quantum numbers
such as spin, isospin, strangeness, linear momenta and orbital momenta.
The intermediate states $B_5B_6$ are those which are allowed by
the Pauli principle, and the energy variable $\omega$ is the starting energy
defined by the single-particle energies
of the incoming external particles $B_3B_4$.
The $G$-matrix is solved using relative and centre-of-mass coordinates,
see e.g., Refs.~\cite{isaac99,sl99} for computational details.
The single-particle energies are given by
\begin{equation}
      \varepsilon_{B_i}=t_{B_i} + u_{B_i} +m_{B_i}
       \label{eq:spenergy}
\end{equation}
where $t_{B_i}$ is the kinetic energy and $m_{B_i}$
the mass of baryon ${B_i}$. The
single-particle potential $u_{B_i}$ is defined by 
\begin{equation}
       u_{B_i}=\mathrm{Re} \sum_{B_j\leq F_j}
       \left\langle B_iB_j\right |
       G(\omega=\varepsilon_{B_j}+\varepsilon_{B_i})
       \left | B_iB_j \right\rangle.
\end{equation}
The linear momentum of the intermediate
single-particle state $B_j$ is limited by the size of the Fermi surface
$F_j$ for particle species $B_j$.
The last equation is displayed in terms of Goldstone diagrams
in Fig.\ \ref{fig:upot}. Diagram (a) represents contributions
from nucleons only as hole states, while diagram (b)
has only hyperons as holes states in case we have a finite hyperon
fraction in $\beta$-stable neutron star matter. The external legs
represent nucleons and hyperons. 

The total non-relativistic energy density, $\varepsilon$, and
the total binding
energy per baryon, ${\cal E}$, can be evaluated from the baryon
single-particle
potentials in the following way
\begin{equation}
\varepsilon=2\sum_{B}
\int_0^{k_F^{(B)}} \frac{d^3 k}{(2\pi)^3}
\left( \frac{\hbar^2k^2}{2M_B}+\frac{1}{2}U_B(k) \right)
\label{eq:binding}
\end{equation}
\begin{equation}
{\cal E}=\frac{\varepsilon}{n} \ ,
\label{eq:binding2}
\end{equation}
where $n$ is the total baryonic density. The density of a given baryon 
species is given by
\begin{equation}
n_{B}=\frac{k_{F_B}^3}{3\pi^2}=x_{B}n \ ,
\end{equation}
where $x_{B}=n_B/n$ is the fraction of
baryon
$B$, which is of course constrained by
\begin{equation}
\sum_{B}x_{B}=1 \ .
\end{equation}

Detailed expressions for the single-particle energies and the $G$-matrices 
involved
can be found in Ref.\ \cite{isaac99}.
In order to satisfy the equations for $\beta$-stable matter summarized
in Eq.\ (\ref{eq:beta_baryonicmatter}), we need to solve
Eqs.\ (\ref{eq:gmatrix}) and (\ref{eq:spenergy}) to obtain
the single-particle energies of the
particles involved at the corresponding Fermi momenta.
Typically, for every total
baryonic density $n=n_N+n_Y$, the density of nucleons plus hyperons,
Eqs.\ (\ref{eq:gmatrix}) and (\ref{eq:spenergy}) were 
solved for five nucleon
fractions and five hyperons fractions and,  for every nucleon and hyperon
fraction, we computed
three proton fractions and three fractions for the relevant hyperons.
The set of equations in Eq.\  (\ref{eq:beta_baryonicmatter}) were
then solved by interpolating between different nucleon and hyperon
fractions.

The many-body approach outlined above is the lowest-order
Brueckner-Hartree-Fock (BHF) method extended to the hyperon sector.
This means also that we consider only
two-body interactions. However, it is well-known from studies of nuclear
matter and neutron star matter with nucleonic degrees of freedom only
that three-body forces are important in 
order to reproduce the saturation 
properties of nuclear matter, see e.g., Ref.\ \cite{apr98} for the most 
recent
approach.  In order to include such effects, we replace the contributions
to the proton and neutron self-energies arising from intermediate
nucleonic states only, see diagram (a) of Fig.\ \ref{fig:upot},
with those derived from Ref.\ \cite{apr98} (hereafter APR98)
where the Argonne $V_{18}$ nucleon-nucleon interaction \cite{v18} 
is used with
relativistic boost corrections and a fitted three-body interaction,
model.
The calculations of Ref.\ \cite{apr98} represent at present perhaps
the most sophisticated many-body approach to dense matter.
In the discussions below we will thus present two sets of results for
$\beta$-stable matter, one where the nucleonic contributions
to the self-energy of nucleons is derived from the baryon-baryon potential
model of Stoks  and Rijken \cite{sr99} and one where the nucleonic 
contributions
are replaced with the results from Ref.\ \cite{apr98} following the 
parametrization
discussed in Eq. (49) of  Ref.\ \cite{hh99}. 
Replacing the nucleon-nucleon part of the interaction model of
Ref.\ \cite{sr99} with that from the
$V_{18}$ nucleon-nucleon interaction \cite{v18}, 
does not introduce large differences at the BHF level. However,
the inclusion of three-body forces as done in Ref.\ \cite{apr98} is 
important.
Hyperonic contributions
will however all be calculated with the baryon-baryon interaction of
Stoks  and Rijken \cite{sr99}.

\subsection{$\beta$-stable neutron star matter}

\begin{figure}[hbtp]
   \includegraphics[totalheight=18cm]{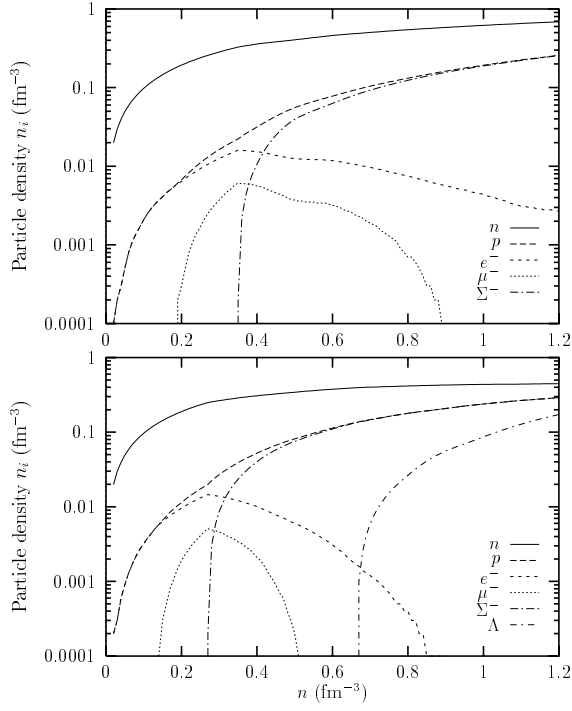}
   \caption{Particle densities in $\beta$-stable neutron star matter
            as functions of the total baryonic density $n$.
            The upper panel represents the results obtained at the
            Brueckner-Hartree-Fock level with the potential of Stoks and
            Rijken \protect\cite{sr99}.
            In the  lower panel the nucleonic part of the self-energy of 
            the nucleons
            has been replaced
            with the EoS of Ref.\ \protect\cite{apr98}. 
            For the latter, (not shown in
            the figure) positrons
            appear at a density 1.18 fm$^{-3}$.}
   \label{fig:fraction}
\end{figure}

\begin{figure}
   \includegraphics[totalheight=20cm]{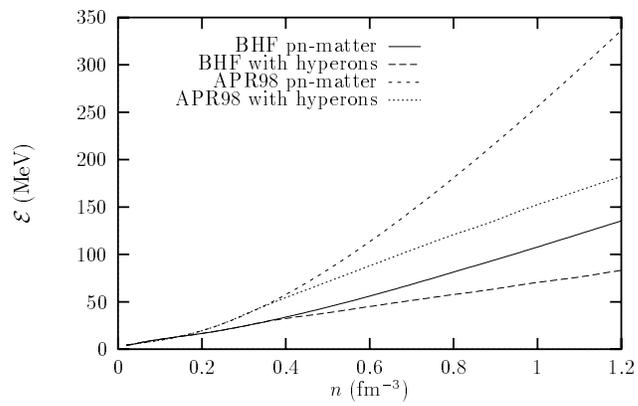}
   \caption{Energy per baryon in $\beta$-stable neutron star matter for
            different approaches  as function of the total baryonic density 
            $n$.
            See text for further details.}
   \label{fig:eosfig}
\end{figure}

The above models for the pure nucleonic part combined with the hyperon
contribution yield
the composition of $\beta$-stable matter, 
up to total baryonic number density 
$n=1.2$ fm$^{-3}$, shown in Fig.\
\ref{fig:fraction}.
The corresponding energies per baryon 
are shown in Fig.\ \ref{fig:eosfig} for both pure nucleonic
(BHF and APR98 pn-matter)
and hyperonic matter (BHF and APR98 with hyperons) in $\beta$-equilibrium
for the same baryonic densities as in Fig.\ \ref{fig:fraction}.

For both types of calculations $\Sigma^-$ appears at densities 
$\sim 2-3 n_0$.
Since the  EoS of APR98 for nucleonic matter yields a stiffer EoS than the
corresponding BHF calculation, $\Sigma^-$ appears at
$n=0.27$ fm$^{-3}$ for the APR98 EoS and $n=0.35$ fm$^{-3}$ for the BHF EoS.
These results are in fair agreement with results obtained from
mean field calculations, see e.g., Refs.\
\cite{prakash97,pke95,ms96}. The introduction
of hyperons leads to a considerable softening of the EoS.
Moreover, as soon as hyperons appear, the leptons tend to disappear,
totally in the APR98 case whereas in the BHF calculation only muons 
disappear.
For the APR98 case, positrons appear at higher densities, i.e., $n=1.18$ 
fm$^{-3}$.  
This result is related to the fact that $\Lambda$ does not appear
at the densities considered here for the BHF EoS.
For the APR98 EoS, $\Lambda$ appears at a density $n=0.67$ fm$^{-3}$.
Recalling $\mu_{\Lambda} = \mu_n = \mu_p + \mu_e$ and that the 
APR98 EoS is stiffer
due to the inclusion of three-body forces, this
clearly enhances the possibility of creating a $\Lambda$ with the APR98 
EoS.
However, the fact that
$\Lambda$ does not appear in the BHF calculation can also, in 
addition to the softer
EoS,
be retraced to a delicate balance between
the nucleonic and hyperonic hole state contributions
(and thereby to features of the baryon-baryon interaction)
to the self-energy of the baryons considered here, see diagrams (a) and (b)
in Fig.\ \ref{fig:upot}. Stated differently, the contributions from 
$\Sigma^-$,
proton and neutron
hole states to the $\Lambda$ chemical potential are not attractive enough
to lower the chemical potential of the $\Lambda$ so that it equals
that of the neutron. Furthermore, the chemical potential of the neutron
does not increase enough since contributions
from $\Sigma^-$ hole states to the neutron self-energy are attractive, 
see e.g., Ref.\ \cite{isaac99} for a detailed account of these
aspects of the interaction model.

\begin{figure}[hbtp]
   \includegraphics[totalheight=18cm]{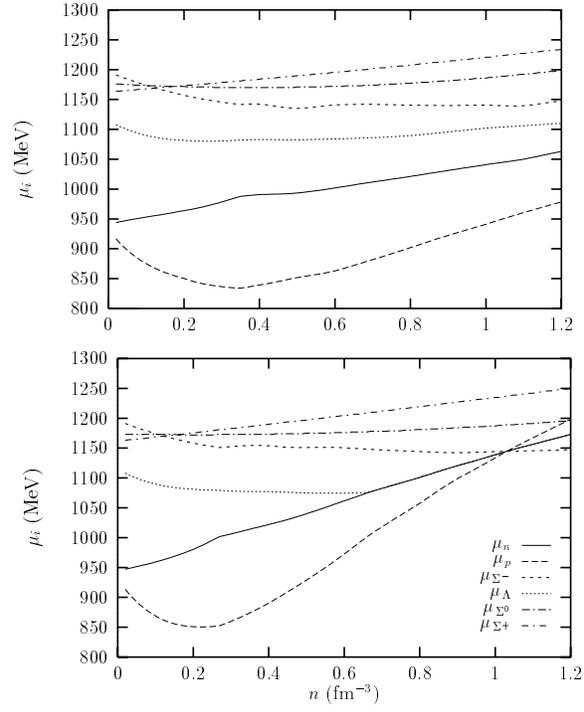}
   \caption{Chemical potentials in $\beta$-stable neutron star matter
            as functions of the total baryonic density $n$.
            The upper panel represents the results obtained at the
            Brueckner-Hartree-Fock level with the potential of Stoks and
            Rijken \protect\cite{sr99}.
            The lower panel includes results obtained
            with the EoS of Ref.\  \protect\cite{apr98}. }
   \label{fig:chempots}
\end{figure}

We illustrate the role played by the two different
choices for nucleonic EoS in
Fig.\ \ref{fig:chempots} in terms of the chemical potentials for 
various baryons
for matter in $\beta$-equilibrium.
We also note that, using the criteria in Eq.\ (\ref{eq:beta_baryonicmatter}),
neither the $\Sigma^0$ nor $\Sigma^+$
do appear for both the BHF and the APR98 equations of state. 
This is due to the
fact that none of the $\Sigma^0$-baryon and  $\Sigma^+$-baryon interactions
are attractive enough. A similar argument  applies to $\Xi^0$ and $\Xi^-$. 
In the latter
case the mass of the particle is $\sim 1315$ MeV and almost 
$200$ MeV in attraction
is needed in order to fullfil e.g., the condition
$\mu_{\Lambda} = \mu_{\Xi^0}= \mu_n$.
This has also been checked by us \cite{selfcascade} in studies of   
the self-energy of $\Xi^-$
in finite nuclei, using the recipe outlined
in Ref.\ \cite{isaacnpa98}. For both light and medium heavy nuclei,
$\Xi^-$ is unbound with the present hyperon-hyperon
interactions, except for version NSC97f of Ref.\ \cite{sr99}. The latter
results in a weakly bound $\Xi^-$, in agreement with the recent studies of
Batty {\em et al.} \cite{batty99}. 
 From the bottom panel of Fig.\ \ref{fig:chempots} we see 
however that $\Sigma^0$
could appear at densities close to $1.2$ fm$^{-3}$.
Thus, for the present densities, which would be within the range of energies
for where the interaction model has been fitted, the only hyperons
which can appear are $\Sigma^-$ and $\Lambda$. 

In summary, using the realistic EoS of Akmal {\em et al.} \cite{apr98} for 
the
nucleonic sector and including hyperons through the 
most recent model for the
baryon-baryon interaction of the Nijmegen group \cite{sr99}, we find
through a many-body calculation for matter in $\beta$-equilibrium that
$\Sigma^-$ appears at a density of $n=0.27$ fm$^{-3}$ while
$\Lambda$ appears at $n=0.67$ fm$^{-3}$.
Due to the formation
of hyperons, the matter is deleptonized
at a density of $n=0.85$ fm$^{-3}$. Within our many-body approach,
no other hyperons appear at densities below $n=1.2$ fm$^{-3}$. 
Although the EoS of Akmal {\em et al.} \cite{apr98} 
may be viewed as the currently
most realistic approach to the nucleonic EoS, our
results have to be gauged with the uncertainty in
the hyperon-hyperon and nucleon-hyperon
interactions. Especially, if the hyperon-hyperon interactions tend to be
more attractive, this may lead to the formation of 
hyperons such as the $\Lambda$,
$\Sigma^0$, $\Sigma^+$, $\Xi^-$ and $\Xi^0$ at lower densities.
The hyperon-hyperon interaction and the stiffness of 
the nucleonic contribution
play crucial roles in the formation of various hyperons. 
These results differ
from present mean field calculations \cite{prakash97,pke95,ms96},
where all kinds of hyperons can appear at the densities considered here.

\subsection{Baryon superfluidity in $\beta$-stable matter}\label{sec:sec3}

A generic feature of fermion systems with attractive interactions is that 
they may be superfluid in a region of the density-temperature plane.   
The $^1S_0$ wave of the nucleon-nucleon interaction is the best known 
and most investigated case in neutron stars, and the results indicate  
that one may expect a neutron superfluid in the inner crust of the 
star and a proton superfluid in the quantum liquid interior, both 
with energy gaps of the order of $1\;{\rm MeV}$ 
\cite{baldo90,baldo92,wambach93,schulze96,khodel96,elgar96}.  
Furthermore, neutrons in the quantum liquid interior may form a 
superfluid due to the attractive $^3P_2$-$^3F_2$ wave of the 
nucleon-nucleon interaction \cite{baldo98}.   
Baryon superfluidity has important consequences for a number of 
neutron star phenomena, including glitches \cite{anderson75} and 
cooling \cite{elgar296}.  If hyperons appear in neutron stars, 
they may also form superfluids if their interactions are sufficiently 
attractive.  The case of $\Lambda$ superfluidity has been investigated 
by Balberg and Barnea \cite{balberg98} using parametrized effective 
$\Lambda$-$\Lambda$ interactions.    Results for $\Lambda$ and 
$\Sigma^-$-pairing using bare hyperon-hyperon interaction models have 
recently been presented by Takatsuka and Tamagaki \cite{taka99}.  
The result of both groups indicate the presence of a $\Lambda$ 
superfluid for baryon densities in the range of $2$--$4n_0$.    
The latter authors also suggest that the formation of a $\Sigma^-$ 
superfluid may be more likely than $\Lambda$-superfluidity.   
Along the lines followed by these authors we will here present 
results for hyperon superfluidity within our model.   

The crucial quantity in determining the onset of superfluidity is the 
energy gap function $\Delta({\bf k})$.  The value of this function 
at the Fermi surface is proportional to the critical temperature 
of the superfluid, and by determining $\Delta$ we therefore map out 
the region of the density-temperature plane where the superfluid may 
exist.  When the $^1S_0$ interaction is the driving cause of the 
superfluidity, the gap function becomes isotropic and depends on the 
magnitude of ${\bf k}$ only.  It can be determined by solving the 
BCS gap equation 
\begin{equation}
\Delta(k)=-\frac{1}{\pi}\int_0^\infty dk'k'^2 \tilde{V}_{^1S_0}(k,k')
\frac{\Delta(k')}{\sqrt{(\epsilon_{k'}-\mu)^2+\Delta(k')^2}}, 
\label{eq:bcsgap}. 
\end{equation}
In this equation, $\epsilon_k$ is the momentum-dependent 
single particle energy in the medium for the particle species in 
question, $\mu$ is the corresponding chemical potential,   
and $\tilde{V}_{^1S_0}$ is the effective pairing interaction. 
At this point we emphasize that using parametrized effective interactions 
in the gap equation can lead to errors.  The gap equation includes 
diagrams also found in the $G$-matrix, and one therefore needs to 
calculate $\tilde{V}$ systematically from microscopic many-body theory 
to avoid double counting of ladder contributions.  The expansion for 
$\tilde{V}$ can be found in e.g. Migdal \cite{migdal67}, and to 
lowest order $\tilde{V}=V$, the free-space two-particle interaction.  
Higher order terms include contributions from e.g. density- and 
spin-density fluctuations.  In this first exploratory calculation  
we will follow Ref. \cite{taka99} and use the bare hyperon-hyperon 
interaction in Eq. (\ref{eq:bcsgap}).   The relevant hyperon fractions 
and single-particle energies are taken from the BHF calculations described 
earlier in this paper.   Details of the numerical solution of the 
gap equation can be found in Ref. \cite{elgar96}.  

\begin{figure}
   \includegraphics[totalheight=20cm]{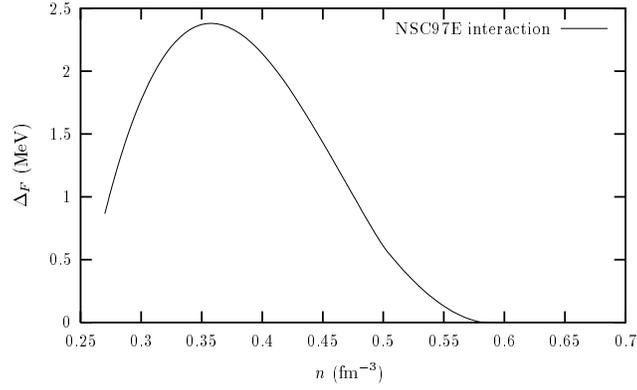}
    \caption{Energy gap $\Delta_F$ as a function of 
             the total baryon density for $\Sigma^-$ hyperons 
             in $\beta$-stable matter.}
    \label{fig:hyperongap}
\end{figure}

Fig. \ref{fig:hyperongap} shows the energy gap $\Delta_F \equiv 
\Delta(k_F^{(\Sigma^-)})$ as a function of 
the total baryon density for 
$\Sigma^-$ hyperons in $\beta$-stable matter for the 
NSC97E model.  Although $\Lambda$ may appear at higher densities, 
the $^1S_0$ 
$\Lambda$-$\Lambda$ matrix elements of the NSC97E interaction 
are all repulsive, and therefore 
the energy gap for $\Lambda$ hyperons would (to lowest order) 
have been zero at all densities, 
i.e. these particles would not have formed a superfluid.     
This is it at variance with the results of Ref. \cite{balberg98}, 
however, as remarked earlier this work employs an effective, 
parametrized interaction to drive the gap equation  and therefore 
overestimates the $\Lambda$ energy gap.  Our $\Sigma^-$ results 
are comparable to those of Ref. \cite{taka99} which were obtained 
with a gaussian soft core parametrization of the bare 
$\Sigma^-$-$\Sigma^-$ interaction.  

If taken at face value these results have implications for neutron 
star cooling.  Since at low densities 
$\Sigma^-$ is the only hyperon species that is present 
in our calculation, the most important contribution to the 
neutrino cooling rate at such densities comes from the reaction 
$\Sigma^- \rightarrow n + e^- + \overline{\nu}_e$.   
According to Ref. \cite{schaab98} the threshold density for this 
reaction to occur is at around $2.4n_0$.  If the $\Sigma^-$s are 
superfluid with energy gaps similar to what we found here, a 
sizeable reduction of the order of $\exp(-\Delta_F/kT)$ may be 
expected in the reaction rate. If neutron stars were to cool
through direct Urca processes, their surface temperatures would
be barely detectable within less than 100 yr of the star's birth.
This is at askance with present observations. Thus, the  
formation of a hyperon
superfluid will clearly suppress the hyperon direct Urca process 
and cooling will most likely proceed through less efficient processes
and bring the results closer to experimental surface temperatures.

\subsection{Structure of neutron stars}\label{sec:sec4}

We end this section with a discussion on neutron star properties 
with the above equations of state.

The best determined neutron star masses are found in binary pulsars
and all lie in the range $1.35\pm 0.04 M_\odot$ \cite{tc99}
except for the nonrelativistic pulsar PSR J1012+5307
of mass
$M=(2.1\pm 0.8)M_\odot$ \cite{vanParadijs1998}. Several X-ray binary
masses have been measured of which the heaviest are
Vela X-1 with $M=(1.9\pm 0.2)M_\odot$ \cite{Barziv1999}
and Cygnus X-2 with
$M=(1.8\pm 0.4)M_\odot$ \cite{OroszKuulkers1999}.  The recent
discovery of high-frequency brightness oscillations in low-mass X-ray
binaries provides a promising new method for determining masses and
radii of neutron stars, see Ref.\ \cite{miller99}. The
kilohertz quasi-periodic oscillations (QPO) occur in pairs
and are most likely the orbital
frequencies of accreting matter
in Keplerian orbits around neutron stars of mass $M$ and its beat
frequency with the neutron star spin.  According to
Zhang {\em et al.} \cite{zhang98} and   Kaaret {\em et al.}
\cite{kaaret1998} the accretion
can for a few QPO's be tracked to its innermost stable orbit.
For slowly rotating stars the resulting mass is
$M\simeq2.2M_\odot({\mathrm{kHz}}/\nu_{QPO})$.  For example, the
maximum frequency of 1060 Hz upper QPO observed in 4U 1820-30 gives 
$M\simeq 2.25M_\odot$ after correcting for the neutron star
rotation frequency.  If the maximum QPO frequencies of 4U 1608-52
($\nu_{QPO}=1125$ Hz) and 4U 1636-536 ($\nu_{QPO}=1228$ Hz) also
correspond to innermost stable orbits the corresponding masses are
$2.1M_\odot$ and $1.9M_\odot$. 
These constraints give us an upper limit for the mass of the order
of  $M\sim 2.2 M_\odot$ and a lower limit $M\sim 1.35 M_\odot$ and 
restrict thereby severely the EoS for dense matter.

\begin{figure}\begin{center}
   \includegraphics[totalheight=20cm]{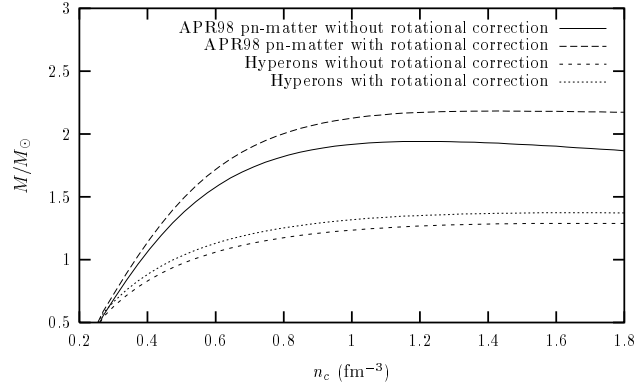}
   \caption{Total mass $M$ for various equations of state. See text for
            further details.}
   \label{fig:mass}
\end{center}\end{figure}

\begin{figure}\begin{center}
   \includegraphics[totalheight=20cm]{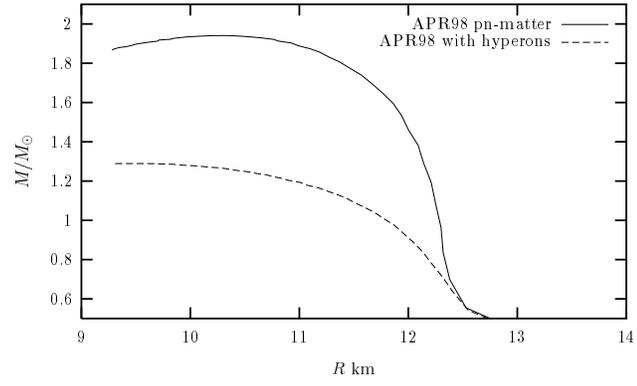}
   \caption{Mass-radius relation without rotational corrections 
            for various various equations of state.}
   \label{fig:massradius}
\end{center}\end{figure}

In the following, we display the results for mass and radius using the
equations of state discussed above.
In order to obtain the radius and mass of a neutron star, we have solved
the Tolman-Oppenheimer-Volkov equation with and without
rotational corrections, following the approach of Hartle \cite{hartle1967},
see also Ref.\ \cite{hh99}.
Our results are shown in 
in Figs.\ \ref{fig:mass} and \ref{fig:massradius}. 
The equations of state we have used are those for
\begin{enumerate}
   \item $\beta$-stable $pn$-matter with the parametrization of the 
         results from Akmal {\em et al.} \cite{apr98} made in
         Ref.\ \cite{hh99}. This EoS is rather stiff compared
         with the EoS obtained with hyperons, see Fig.\ \ref{fig:eosfig}.
         The EoS yields a maximum mass $M\sim 1.9M_\odot$ without
         rotational corrections and $M\sim 2.1M_\odot$ when rotational
         corrections are included. The results for the mass are shown
         in Fig.\ \ref{fig:mass} as functions of central density $n_c$.
         They are labelled as $pn$-matter with and without rotational
         corrections. The corresponding 
         mass-radius relation (without rotational corrections)  is shown
         in Fig.\ \ref{fig:massradius}.
    \item The other EoS employed is that which combines the nucleonic
          part of Ref.\ \cite{apr98} with the computed hyperon 
          contribution. As can be seen from Fig.\ \ref{fig:mass}, the softening
          of the EoS due to additional binding from hyperons leads to a
          reduction of the total mass. Without rotational corrections, we obtain
          a maximum mass $M\sim 1.3M_\odot$ whilst the rotational correction 
          increases the mass to  $M\sim 1.4M_\odot$. The size of the 
          reduction, $\Delta M\sim 0.6-0.7M_\odot$, and the obtained 
           neutron star masses due to hyperons are comparable
          to those reported by Balberg {\em et al.} \cite{blc99}. 
\end{enumerate}

\begin{figure}\begin{center}
   \includegraphics[totalheight=20cm]{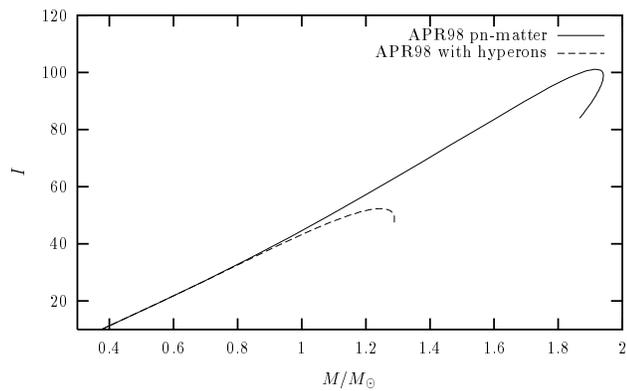}
\caption{The maximum moment of Inertia $I$, in units of $M_{\odot}$km$^2$.
         Same equations of state as in the preceeding figure.
         All results are for $\beta$-stable matter and rotational
         corrections have not been included in the total mass.}         
   \label{fig:massi}
\end{center}\end{figure}

There are other features as well to be noted from Fig.\ \ref{fig:mass}.
The EoS with hyperons reaches a maximum mass at a central density 
$n_c\sim 1.2-1.3$ fm$^{-3}$. In Fig.\ \ref{fig:fraction} 
we showed that the only hyperons which can appear at these densities
are $\Lambda$ and $\Sigma^-$.
If other hyperons were to appear at higher
densities, this would most likely lead to a further softening of the EoS, and 
thereby smaller neutron star masses. Furthermore, the softer EoS yields
also a smaller moment of inertia, as seen in Fig.\ \ref{fig:massi}.

The reader should however note that our calculation of hyperon
degrees freedom is based on a non-relativistic Brueckner-Hartree-Fock
approach. Although the nucleonic part extracted from Ref.\ \cite{apr98},
including three-body forces and relativistic boost corrections, is to be 
considered as a benchmark calculation for nucleonic degrees of freedom,
relativistic effects in the hyperonic calculation could result in a
stiffer EoS and thereby larger mass. However, relativistic mean 
field calculations
with parameters which result in a similar composition of matter as
shown in Fig.\ \ref{fig:fraction}, result in similar masses
as  those reported in Fig.\ \ref{fig:mass}.
In this sense, our results may provide a lower and upper bounds for the 
maximum mass. This  leaves two natural options
when compared to the observed neutron star masses.
If the above heavy neutron stars prove erroneous 
by more detailed observations
and only masses like those of binary pulsars are found, 
this may indicate that heavier neutron stars simply are not
stable which in turn implies a soft EoS, or that a
significant phase transition must occur already at a few times nuclear
saturation densities. Our EoS with hyperons would fit into this case, although
the mass without rotational corrections is on the lower side.
Else, if the large masses from QPO's are confirmed, then the EoS
for baryonic matter needs to be stiffer and in our case, this would
rule out the presence of hyperons up to densities 
$\sim 10n_0=1.2$ fm$^{-3}$.

Although we have only considered the formation of hyperons in neutron
stars, transitions to other degrees of freedom such as quark matter,
kaon condensation and pion condensation may or may not take place
in neutron star matter.
We would however like to emphasize that the hyperon formation mechanisms
is perhaps the most robust one and is likely to occur in the interior
of a neutron star, unless the hyperon self-energies are strongly repulsive 
due
to repulsive hyperon-nucleon and hyperon-hyperon interactions, a repulsion
which  would contradict
present data on hypernuclei \cite{bando}.
The EoS with hyperons yields however neutron star masses without
rotational corrections which are even below $\sim 1.4M_\odot$.
This means that our EoS with hyperons needs to be stiffer, 
a fact which may in turn
imply that more complicated many-body terms not included in our 
calculations,
such as three-body forces between nucleons and hyperons and/or relativistic
effects,  are needed.

\section{Conclusions}\label{sec:sec5}

Employing the recent 
parametrization
of the free baryon-baryon potentials
for the complete  baryon octet
of Stoks and Rijken \cite{sr99}, we have performed a
microscopic many-body calculation of the structure of
$\beta$-stable neutron star matter including 
hyperonic degrees of freedom. 
The potential model employed allows for the presence of only two
types of hyperons up to densities ten times nuclear matter
saturation density. These hyperons are
$\Sigma^-$ and $\Lambda$. The interactions for strangeness
$S=-1$, $S=-2$, $S=-3$ and $S=-4$ are not attractive enough to allow
the formation of other hyperons. The presence of hyperons leads however
to a considerable softening of the EoS, entailing a corresponding reduction
of the maximum mass of the neutron star. With hyperons, we obtain maximum
masses of the order $M\sim 1.3-1.4 M_\odot$.

In addition, since $\Sigma^-$ hyperons appear already at total baryonic 
densities
$\sim n=0.27$ fm$^{-3}$), we have also considered the possibility of forming
a hyperon superfluid. The latter would in turn quench the 
increased emission of neutrinos due to the presence of hyperons.
Within our many-body approach, we find that 
$\Sigma^-$ forms a superfluid in the $^1S_0$ wave, whereas the
$\Lambda-\Lambda$ interaction for the same partial wave leads to a vanishing
gap for the potential model employed here.

We are much indebted to H.~Heiselberg, H.-J.~Schulze 
and V.~G.~J.~Stoks for many usuful comments.
This work has been supported by the DGICYT (Spain) Grant PB95-1249 and
the Program SCR98-11 from the Generalitat de Catalunya. One of the authors
(I.V.) wishes to acknowledge support from a doctoral fellowship of the
Ministerio de Educaci\'on y Cultura (Spain).




\end{document}